# Title: Programming Correlated Magnetic States via Gate Controlled Moiré Geometry

**Authors:** Eric Anderson[1], Feng-Ren Fan[2,3], Jiaqi Cai[1], William Holtzmann[1], Takashi Taniguchi[4], Kenji Watanabe[5], Di Xiao[6,1], Wang Yao[2,3*], and Xiaodong Xu[1,6*]

**Affiliations:**

[1]Department of Physics, University of Washington; Seattle, WA, USA.

[2]Department of Physics, University of Hong Kong; Hong Kong, China.

[3]HKU-UCAS Joint Institute of Theoretical and Computational Physics at Hong Kong; Hong Kong, China.

[4]International Center for Materials Nanoarchitectonics, National Institute for Materials Science; Tsukuba, Ibaraki 305-0044, Japan.

[5]Research Center for Functional Materials, National Institute for Materials Science; Tsukuba, Ibaraki 305-0044, Japan.

[6]Department of Materials Science and Engineering, University of Washington; Seattle, WA, USA.

[*]**Corresponding Authors**. wangyao@hku.hk; xuxd@uw.edu

**Abstract:** Understanding quantum many-body systems is at the heart of condensed matter physics. The ability to control the underlying lattice geometry of a system, and thus its many-body interactions, would enable the realization of and transition between emergent quantum ground states. Here, we report in-situ gate switching between honeycomb and triangular lattice geometries of an electron many-body Hamiltonian in R-stacked MoTe2 moiré bilayers, resulting in switchable magnetic exchange interactions. At zero electric field, we observe a correlated ferromagnetic insulator near one hole per moiré unit cell (v=-1), i.e., a quarter-filled honeycomb lattice, with a widely tunable Curie temperature up to 14K. Fully polarizing layer pseudospin via electric field switches the system into a half-filled triangular lattice with antiferromagnetic interactions. Further doping this layer-polarized superlattice introduces carriers into the empty layer, tuning the antiferromagnetic exchange interaction back to ferromagnetic. Our work demonstrates R-stacked MoTe2 moirés to be a new laboratory for engineering correlated states with nontrivial topology.

**Main Text:**

The physical properties of crystalline solids are fundamentally determined by their lattice structure. The ability to control lattice parameters would thus enable access to a complex electronic phase diagram. Moiré superlattices of two-dimensional van der Waals crystals have recently emerged as powerful synthetic quantum materials capable of achieving designer Hamiltonians with controllable superlattice constants, layer stacking arrangements, and Coulomb interaction strengths (*1*). So far, a plethora of correlated and topological electronic states have been demonstrated in the triangular lattice (*2–11*). However, the honeycomb lattice, a model system for investigating strongly correlated phenomena, remains to be explored. In addition, *in situ* controllable superlattice geometry has not been realized, which would be a fundamentally new approach to electrically tune phase transitions between states with distinct symmetries.

Rhombohedral (R-) stacked transitional metal dichalcogenide (TMD) moiré bilayers may offer such opportunities (*12–17*). As shown in Fig. 1A, the moiré potential has two degenerate energy minima within a supercell, the MX (B sublattice) and XM (C sublattice) sites. Here, MX denotes the transition metal atoms (M) of one layer sitting atop of chalcogen atoms (X) of the other. The corresponding moiré orbitals are localized in opposite layers, forming a honeycomb lattice with the sublattice pseudospin locked to the layer pseudospin (Fig. 1B). In this picture, the application of a vertical electric field induces layer polarization and breaks energy degeneracy between the moiré orbitals localized in the B and C sublattices, leading to a transition from honeycomb to triangular lattice symmetry (Fig. 1B).

The R-stacked homobilayer moiré has been theoretically predicted to host an array of intriguing phenomena, including the quantum spin Hall effect, ferroelectric Mott insulators, integer and fractional quantum anomalous Hall states, and electric field induced electronic phase transitions (*13*, *14*, *18–25*). Experimentally, moiré ferroelectricity (*26*), interlayer exciton-electric polarization coupling (*27*), correlated electronic phases (*28*, *29*), and quantum criticality near one hole per moiré unit cell ($v$ = -1) (*30*) have been reported in twisted R-stacked bilayers. Here, we show that R-stacked twisted bilayer $MoTe_2$ is a new model system for exploring interaction induced magnetism with electrically tunable moiré geometry.

We fabricate near 4-degree twisted $MoTe_2$ bilayers in a dual gated structure, which allows independent control of carrier density $n$ and displacement field $D$. There is a small built-in displacement field of 0.04 V/nm in the device, which is likely caused by imperfect sample fabrication. For simplicity, an effective $D$ with this built-in field offset subtracted is presented in the rest of the paper (see Methods in Supplementary Materials). To characterize the device properties, we perform photoluminescence (PL) (Fig. 1C) and optical reflectance (fig. S1) measurements versus $n$ at fixed $D$ = 0 V/nm. The experimental temperature is 1.6 K unless otherwise specified. Comparing this twisted bilayer PL with doping dependent monolayer data taken under identical conditions (Fig. 1D), we observe similar general behavior. Namely, there is a charge neutral exciton near zero gate voltage and trions which appear as gating induces electrostatic doping (*31*).

The similarity of the monolayer and bilayer doping dependent PL data suggests that twisted $MoTe_2$ bilayer is a direct bandgap semiconductor. Similar direct bandgap nature has been observed in as-exfoliated 2H stacked bilayer $MoTe_2$ (*32*, *33*) where band-edges are at the corners of the hexagonal Brillouin zone (or valleys). The direct-bandgap property is distinct from other TMD

homo- and hetero-bilayers, which have an indirect bandgap. Therefore, in twisted MoTe$_2$ bilayers, the valley composition is unambiguous for both charged carriers and excitons. The latter is particularly appealing as a sensitive optical probe of correlated physics, due to the sharpness of the exciton spectra (~ 5 meV, see fig. S2).

Distinct from the monolayer, however, the trion photoluminescence in twisted bilayer is strongly modulated by doping. This observation resembles the behavior of the interlayer exciton in WS$_2$/WSe$_2$ heterobilayer, which has been used to probe the formation of correlated electronic states at integer and fractional filling in moiré superlattices (*8, 34–36*). By comparing both PL and reflectance data, we can identify the integer filling factor (*v* =1, 2, -1) and infer a twist angle of about 3.9$^0$ (Supplementary Section 1.3). A significant feature of the doping dependent photoluminescence is the drop in trion intensity at integer fillings |*v*|=1 and 2. This supports the interpretation of insulating state formation, as this would reduce the free carrier population needed to form trions, thus suppressing the trion PL.

We perform reflective magnetic circular dichroism (RMCD) measurements to investigate magnetic interactions. The optical excitation is chosen at 1.12 eV with a bandwidth of 30 meV (see fig. S3 for wavelength dependence). Fig. 1E shows the RMCD signal as a function of *v* and *D*. The data is taken by first initializing the sample at out-of-plane magnetic field $\mu_o$H =0.5T and then sweeping back to zero field. Remnant RMCD signal, the signature of a ferromagnetic state, is pronounced on the hole side centered around *v*=-1. As we demonstrate below, this is a two-dimensional magnetic phase diagram over *v* and *D*. The focus of this work is to understand the phase diagram.

We start with investigation of the magnetic state at *D*=0, i.e., in the honeycomb lattice. Fig. 2A shows the RMCD signal at *v*=-1 versus out-of-plane $\mu_o$H swept down and up. A pronounced hysteresis loop with a width of about 40 mT is observed, a hallmark of ferromagnetism. Sharp switching of the RMCD signal near the critical field implies a spin-flip transition with an out-of-plane easy axis. Fig. 2B is the temperature dependence of RMCD vs $\mu_o$H, further confirming the ferromagnetic state. Both the hysteresis loop width and remnant RMCD signal decrease as temperature increases, and eventually vanish above the Curie temperature (T$_C$) of about 14K.

The observed ferromagnetic state near *v*=-1 is consistent with the ground state of a quarter filled honeycomb lattice, gapped by the next-nearest neighbor complex hopping in twisted R-type homobilayers (*14*). The spin-resolved band structure is obtained from Hartree-Fock calculations (see Methods). As shown in Fig. 2C, a fully spin polarized valence band is observed. Calculation shows that at *D*=0, hole density is equally distributed between the B sublattice in one layer and C sublattice in the other layer (Fig. 2D). The B and C orbitals have appreciable spatial overlap where the carrier wavefunction becomes layer hybridized, such as at the A corners of the moiré unit cell and the middle points between B and C. It is the Coulomb exchange through the spatial overlap that gives rise to nearest neighbor ferromagnetic interactions. The calculation also finds that the spin band is topological with a Chern number of 1, i.e., *v*=-1 is a quantum anomalous Hall (QAH) insulator. This result is consistent with theoretical predictions that a QAH state can be realized in a quarter filled honeycomb lattice (*37*) such as graphene. The existence of a QAH state is also predicted in twisted MoTe$_2$ bilayers under similar conditions (*13, 21*).

The ferromagnetic state can be tuned by electrostatic doping. The top and bottom panels of Fig. 2E are the RMCD intensity plots versus *v*, for $\mu_o$H swept down (top panel) and up (bottom panel), respectively (see fig. S4 for linecuts). The difference of the two plots yields the residual RMCD signal and hysteresis loop width. As shown in Fig. 2F, the ferromagnetic state exists for the range of $v$ between -0.5 and -1.3, indicating a possible transition from ferromagnetic insulator at $v$=-1 to ferromagnetic metal upon doping. The hysteresis loop width appears jumpy over changing $v$ and shrinks near the phase space boundary. The jumpiness is likely due to domain dynamics near the spin flip transition. We performed temperature dependent RMCD at selected filling factors within this $v$ range (fig. S5). The extracted $T_C$ varies by a factor of four, from about 14K near $v$=-1 to 3K (limited by our base temperature of 1.6K) near the phase boundary (Fig. 2G). This result demonstrates strongly doping dependent ferromagnetic properties.

We measure RMCD signal as a function of $D$ at $v$ = -1. Figures 3A&B are the RMCD intensity plots versus $\mu_o$H swept down and up, respectively. The difference between them (Fig. 3C) demonstrates the remanent RMCD signal and hysteretic effects, highlighting the electric field tunability of the ferromagnetic state. The hysteresis loop width remains finite as $D$ increases, until the hysteretic behavior finally vanishes for $D$ values larger than ~0.2 V/nm. The Curie temperature is correspondingly tuned upon increasing $D$ by a factor of four (Fig. 3D). The suppression of the ferromagnetic interaction is due to the charge redistribution between the layers as $D$ field increases. Above the critical displacement field value, full layer polarization develops (see fig. S6). At this value of D, the on-site energy difference between the two honeycomb sublattices due to the displacement field is sufficiently large that all carriers become confined to a single layer sublattice, with the sublattice of the opposite layer empty. In this regime, Coulomb exchange between the filled sites (either B or C, depending on the direction of the applied D field) is quenched by their larger separation compared to the honeycomb lattice. The system is reminiscent of the extensively discussed triangular lattice Hubbard model in heterobilayer TMD moirés ($2$–$7$).

To reveal the magnetic interactions in the fully layer pseudospin polarized triangular lattice, we perform RMCD measurements versus $\mu_o$H as a function of temperature. Fig. 3E shows the results at $D$=0.32 V/nm. A paramagnetic-like response curve with saturated RMCD signal at high magnetic field is observed at base temperature, and disappears within the applied magnetic field range at high temperature. We extract the slope of the RMCD signal curve near $\mu_o$H=0 (simplified as $\frac{\partial \mathcal{R}}{\partial H}|_{H=0}$ for use as a proxy of the magnetic susceptibility $\chi$, as in previous reports ($38$). We then plot 1/slope versus temperature and fit the data to the Curie-Weiss law, $\chi = \frac{c}{T-\theta_c}$ (Fig. 3F). The fitted line intercepts the temperature axis at a negative value, yielding a Curie-Weiss temperature $\theta_c$ of about -3.5K. This negative $\theta_c$ demonstrates an antiferromagnetic interaction between local moments in the moiré traps. This inferred antiferromagnetic interaction is consistent with the 120° Néel order of a half-filled triangular lattice with strong onsite $U$.

To explore the nature of the $D$ field induced magnetic phase transition, we performed RMCD measurements for $D$ swept forward and backward at $\mu_o$H=0. The data is taken in the slightly under-doped regime ($v$=-0.9) to avoid the complications due to domain effects near $v$=-1 (see fig. S7). As shown in Fig. 3G, there is no appreciable hysteresis between the curves near the phase transition boundary. As hysteresis is a signature of a first order phase transition, the lack of hysteresis implies that the D field induced magnetic phase transition is second order. To further support this understanding, we extract the RMCD slope near $\mu_o$H=0 at selected $D$ values in the fully layer polarized state, i.e., in the antiferromagnetic interaction regime. The extracted RMCD slopes $\frac{\partial \mathcal{R}}{\partial H}$, proportional to the magnetic susceptibility $\chi$, rapidly increase as D approaches the phase transition boundary. The singular behavior of $\frac{\partial \mathcal{R}}{\partial H}$ near the phase transition is consistent with the expectation of a second order phase transition. In essence, what we have demonstrated is a unique magnetoelectric effect of a correlated charge insulating state by tuning the moiré Hamiltonian from a honeycomb to triangular lattice.

Lastly, we turn to a discussion of the reemergence of ferromagnetic states with increasing hole doping above $v$=-1 at large $D$, as seen in the 2D phase diagram in Fig. 1E. Using $D$=0.26 V/nm as a representative large $D$ value, Fig. 4A shows RMCD measurements at selected $v$ varying from -1 to -1.5. As doping increases away from the $v$=-1 insulating state, RMCD hysteresis signal appears, increases, and eventually vanishes again. Fig. 4B plots the residual RMCD signal versus $v$ between $\mu_0 H$ swept down and up (see fig. S8 for raw data). It highlights the phase space of the revival of the ferromagnetic states. The ferromagnetic phase at $D$=0.26V/nm spans a range of $v$ between about -1.1 and -1.4, a significant shift with a reduced range compared to the $D$=0 ferromagnetic phase centered around $v$=-1.

The above results showcase another unique capability of the twisted MoTe$_2$ bilayer system– controlling the magnetic order of the correlated insulating state in a triangular lattice by proximity to a moiré layer with tunable doping. Figures 4C&D illustrate the idea. Starting at $v$=-1 at large $D$, the top layer is an insulator with antiferromagnetic interactions between local moments on the B sites of the moiré, and the bottom layer is an empty triangular lattice (Fig. 4C). As we increase the doping to $v$=-(1+$x$) with $x$<1, as long as the onsite Columb interaction $U$ is larger than the $D$ field induced charge transfer gap, the extra carrier $x$ goes to the first miniband of the lower layer, while chemical potential remains within the charge gap of the top layer (Fig. 4D). This realizes a Mott insulating state in proximity to a doped moiré layer.

Distinct from the proposed gate tunable exchange interactions of moiré Kondo lattices in hetero-bilayer geometry (*39*, *40*), where the carriers in the conducting layer are not trapped by the moiré potential, the doped layer in our case is a triangular lattice with presumably comparable moiré trapping potential to the other Mott insulator layer. Fig. 4E compares the calculated charge distribution under finite $D$ field at integer filling ($v$=-1) and with additional doping $x$ ($v$=-(1+$x$)). Here, the extra carriers are centered at the B sites of the bottom layer. With this partial filling of the B sublattice, the ferromagnetic Coulomb exchange from the appreciable wavefunction overlap between the nearest-neighbor B and C sites starts to dominate over the antiferromagnetic kinetic exchange between the C sites in the same layer. Thus, ferromagnetic order develops in the bilayer system.

In summary, we have demonstrated that the twisted bilayer MoTe$_2$ in R-stacking provides a new platform for exploring a many-body Hamiltonian with gate tunable lattice geometry and exchange interactions. An immediate opportunity is to explore the rich magnetic phase diagram in the honeycomb lattice, such as the predicted AFM state at v=-2 (half filled honeycomb lattice) and FM state at v=-3 (filling of the second moiré flat band) (*13*). R-stacked MoTe$_2$ also hosts moiré electric polarization with opposite dipole orientation in adjacent moiré orbitals. With this feature, combined with the direct band-gap properties and magnetic states, it is feasible to explore tunable moiré multiferroicity with excitonic spectroscopy (*19*). Theory also predicts a variety of topological states, such as a QAH state with an electrically tunable topological phase transition - an interesting direction for electrical transport measurements (*13*, *14*, *18*, *21*). In addition, the moiré minibands are expected to become flatter as the twist angle is decreased. In fact, bilayer MoTe$_2$ and WSe$_2$ twisted at an angle of about 1.4 degrees has been predicted to host fractional QAH states (*18*, *21*). Therefore, it will be fascinating to engineer and explore magnetic interactions of correlated states at fractionally filled minibands as well as associated topological states in small twist angle bilayer MoTe$_2$. By engineering multilayer MoTe$_2$, such as in a twisted monolayer-bilayer system, it may be possible to realize a platform for investigating a gate tunable moiré Kondo lattice (*39*). Our work paves the way for these possibilities.

**Acknowledgments:** We thank Jiu-Haw Chu and Liang Fu for helpful discussions and Zhaoyu Liu for assistance with sample preparation.

**Funding**: The work is mainly supported by DoE BES under award DE-SC0018171. Sample fabrication was partially performed using instrumentation supported by the U.S. National Science Foundation through the UW Molecular Engineering Materials Center (MEM·C), a Materials Research Science and Engineering Center (DMR-1719797). EA and WH acknowledge the support by the National Science Foundation Graduate Research Fellowship Program under Grant No. DGE-2140004. Work at HKU is supported by the Research Grants Council of Hong Kong SAR (AoE/P-701/20, HKU SRFS2122-7S05). WY acknowledges support from Tencent Foundation. K.W. and T.T. acknowledge support from the JSPS KAKENHI (Grant Numbers 19H05790, 20H00354 and 21H05233). XX acknowledges support from the State of Washington funded Clean Energy Institute and from the Boeing Distinguished Professorship in Physics.

**Author contributions:** XX and WY conceived the project. EA fabricated and characterized the samples, assisted by JC. EA performed the magneto-optical measurements, assisted by JC and WH. EA, XX, DX, FF, WY analyzed and interpreted the results. FF and WY performed electronic structure calculations. TT and KW synthesized the hBN crystals. EA, XX, FF, WY wrote the paper with input from all authors. All authors discussed the results.

**Competing interests:** The authors declare no competing financial interests.

**Data and materials availability:** The datasets generated during and/or analyzed during this study are available from the corresponding author upon reasonable request.

**Supplementary Materials:**

Materials and Methods

Supplementary Text

Figs. S1 to S8

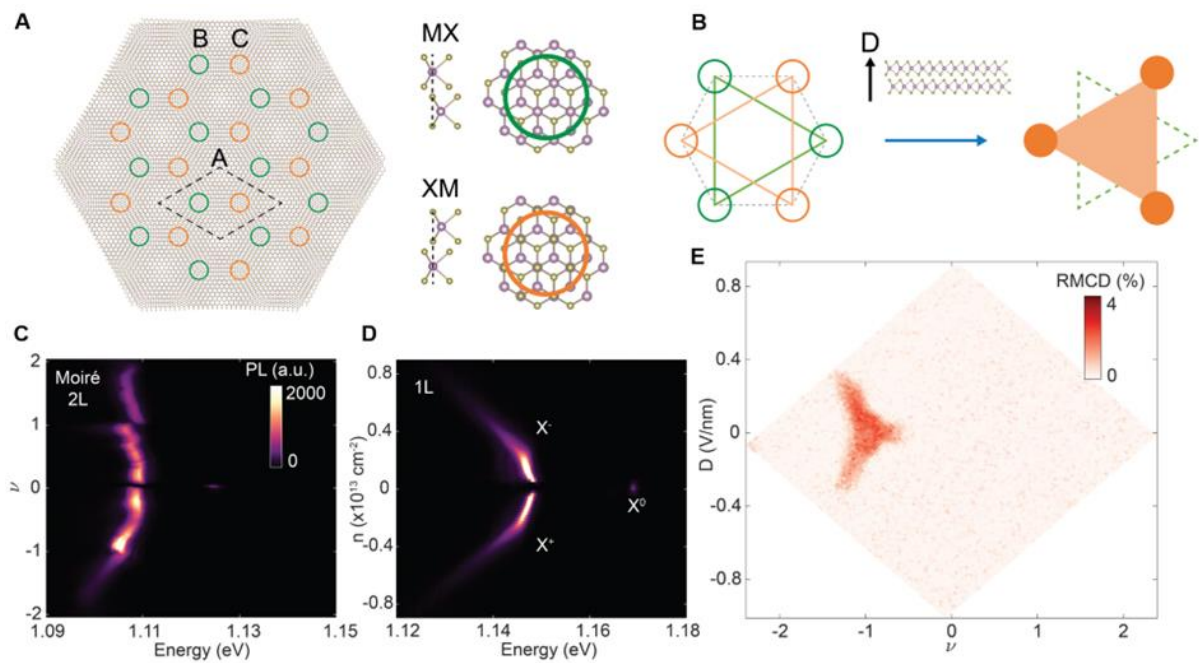

**Fig. 1. Gate tunable moiré geometry, correlated states, and magnetic response in twisted MoTe₂ bilayer.** (**A**) Cartoons of small angle twisted MoTe$_2$ bilayer in rhombohedral stacking. High symmetry points with local energy minima are highlighted. The green circles correspond to MX sites (B sublattice), where the metal atom M in the top layer is aligned with the chalcogen atom X in the bottom layer. Orange circles are the corresponding XM sites (C sublattice). Dotted lines denote a single moiré unit cell. (**B**) Application of vertical electrical field ($D$) lifts the energy degeneracy of the layers and switches the moiré superlattice geometry from honeycomb (left) to triangular (right). (**C**) Photoluminescence intensity plot of the twisted bilayer as a function of moiré filling factor ($v$) and photon energy. (**D**) Doping dependent photoluminescence of monolayer MoTe$_2$, showing exciton (X$^o$) and trion (X$^+$ and X$^-$) features. (**E**) Reflective magnetic circular dichroism (RMCD) signal intensity plot as a function of $v$ and $D$ of the twisted bilayer without magnetic field. The non-zero RMCD signal is observed centered around $v$=-1 and symmetric in $D$, and signifies a two-dimensional ferromagnetic phase diagram.

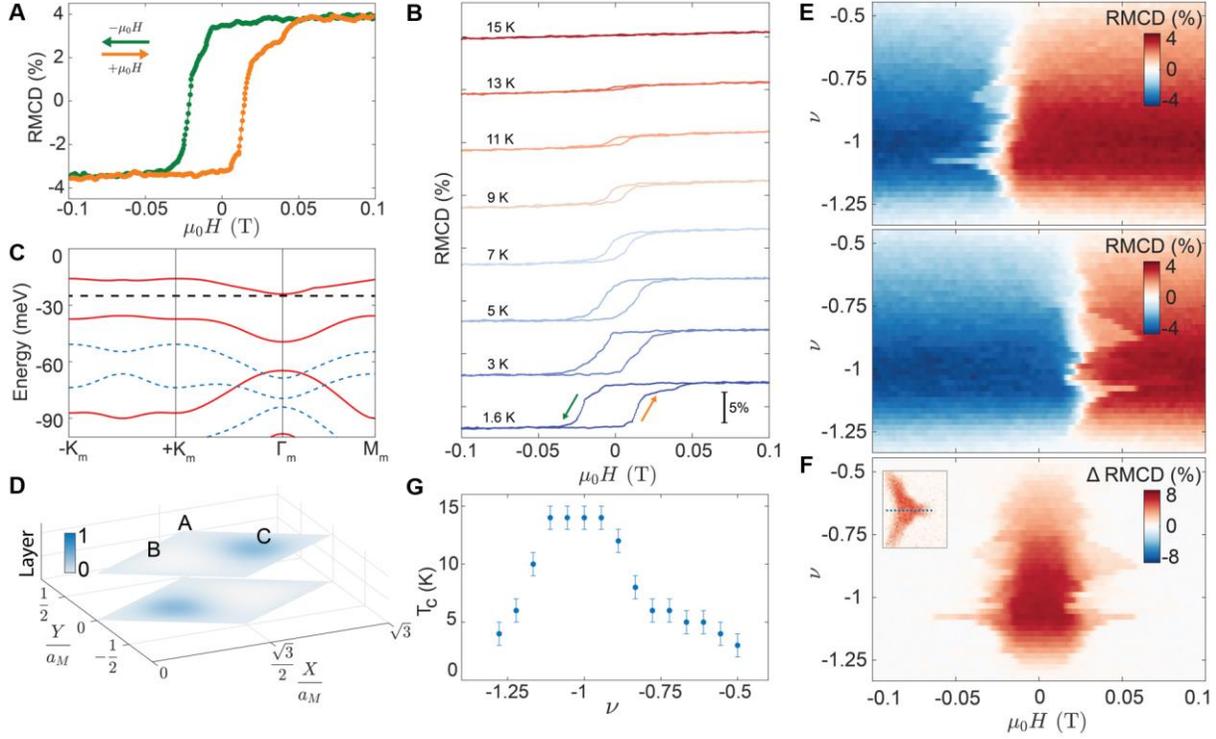

**Fig. 2. Ferromagnetism near quarter-filled honeycomb lattice.** All data are taken at $D$=0. (**A**) RMCD signal versus out-of-plane magnetic field ($\mu_o$H) swept down and up at $\nu$=-1. The observed hysteretic behavior demonstrates the ferromagnetic state. (**B**) Temperature dependent RMCD at $\nu$=-1, showing behavior typical of a ferromagnetic state. Data are offset for clarity. (**C**) Calculated valley-resolved moiré band structure in 3.9$^0$ twisted bilayer MoTe$_2$. Solid red and blue dashed lines represent spin up and down bands, respectively. Black dashed line denotes the chemical potential. (**D**) Calculated hole density spatial distribution over the moiré unit cell at $\nu$=-1. Density is distributed evenly between the bottom layer at the B sites and the top layer at the C sites, with appreciable spatial overlap between the nearest neighbor B and C sites. Color saturation corresponds to normalized hole density. (**E**) RMCD signal intensity plot versus filling factor $\nu$ and $\mu_o$H swept down (top) and up (bottom). (**F**) Difference of the top and bottom panels in (E), giving the hysteretic component of the RMCD. This highlights the filling factor phase space for the ferromagnetic state. Inset: the blue dotted line in the 2D phase diagram corresponds to the range of $\nu$ in (E-F). (**G**) Filling factor dependent Curie temperature. Error bars correspond to the temperature sampling resolution.

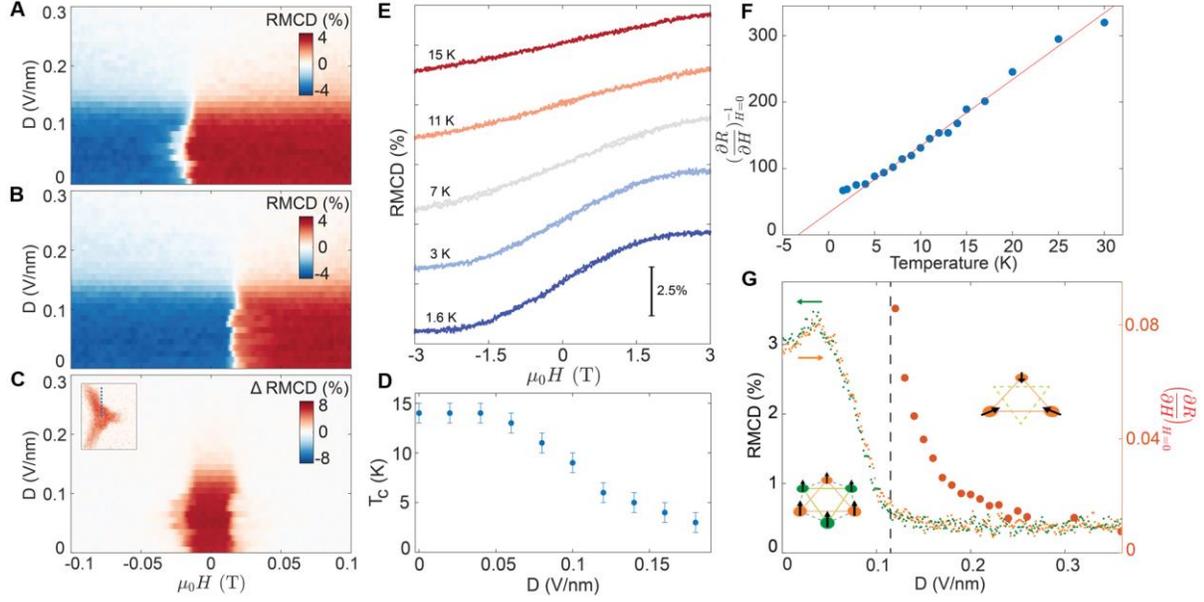

**Fig. 3. Electric field tunable ferromagnetic to antiferromagnetic exchange interactions.** All data are taken at $\nu$=-1, except for data in (G) taken at $\nu$=-0.9. (**A**-**B**) RMCD signal versus $D$ and $\mu_0$H swept down (A) and up (B). (**C**) Hysteretic component of the RMCD versus $D$, extracted from the difference between (A) and (B). The blue dotted line in the inset denotes the phase space of the data taken in (A-C). Ferromagnetic states can be continuously tuned by electric field and eventually switched off once full layer polarization is achieved (i.e., a half-filled triangular lattice). (**D**) $D$ field tunable Curie temperature. (**E**) Temperature dependent RMCD in the fully layer polarized state at $D$=0.32 V/nm. Data are offset for clarity. (**F**) Curie-Weiss fit (red line) of inverse RMCD slope at zero magnetic field ($\left(\frac{\partial \mathcal{R}}{\partial H}\right)_{H=0}^{-1}$) vs temperature. Negative Curie-Weiss temperature $\theta_c$ implies antiferromagnetic interactions between local moments. (**G**) RCMD ($\mu_0$H = 0) versus $D$ swept down (green) and up (orange). Red dots are the extracted RMCD slope $\left(\frac{\partial \mathcal{R}}{\partial H}\right)_{H=0}$ as $D$ approaches the ferromagnetic phase from the antiferromagnetic interaction side, showing singular behavior. Insets denote the spin arrangements with the antiferromagnetic and ferromagnetic interactions in the triangular and honeycomb lattices, respectively.

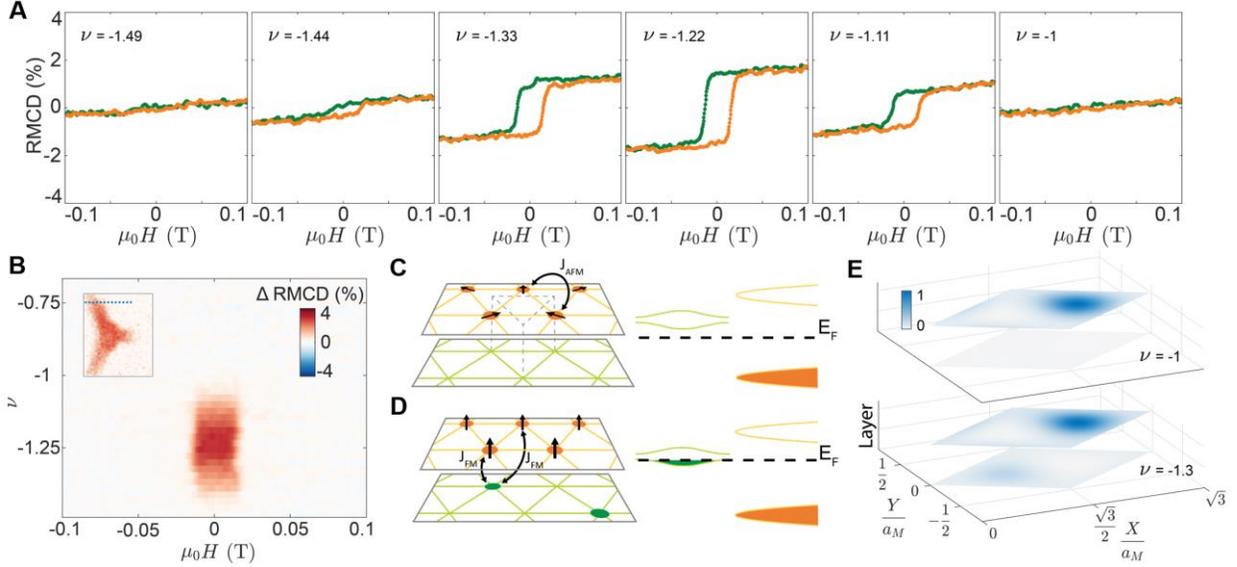

**Fig. 4. Proximity control of the magnetic exchange interaction in the Mott state.** All data are taken at $D$=0.26 V/nm. (**A**) RMCD signal versus μ₀H at selected filling factors. (**B**) Hysteretic component of the RMCD signal versus $v$, highlighting the ferromagnetic state in the phase space of $v$. The data is obtained by taking the difference of the RMCD signal between magnetic field swept down and up (fig. S8). Inset: the blue dotted line in the 2D phase diagram corresponds to phase space where the data is taken. (**C**) Left: Schematic of twisted bilayer MoTe₂ at $v$=-1 with fully polarized layer pseudospin. The top layer is a half-filled triangular lattice with antiferromagnetic interactions between the local moments, while the bottom layer is unfilled. Right: Schematic of band alignment. The chemical potential is in the charge gap of the top layer while the moiré minibands of the bottom layer are empty. (**D**) As in (C) but with filling factor $v$=-(1+$x$). Given that the chemical potential is still in the charge gap of the top layer while the miniband of the bottom layer is partially filled (right panel), the top layer remains in a half-filled Mott state while the bottom layer becomes a doped triangular lattice. The direct Coulomb exchange of the partially filled B orbitals in this doped layer with the fully filled C orbitals in the top layer gives rise to ferromagnetism. (**E**) Calculated charge distribution in the biased condition. Top: $v$=-1 and bottom $v$=-1.3.

# Supplementary Materials for

## Programming Correlated Magnetic States via Gate Controlled Moiré Geometry

Eric Anderson[1], Feng-Ren Fan[2,3], Jiaqi Cai[1], William Holtzmann[1], Takashi Taniguchi[4], Kenji Watanabe[5], Di Xiao[6,1], Wang Yao[2,3*], and Xiaodong Xu[1,6*]

Correspondence to: wangyao@hku.hk; xuxd@uw.edu

**This PDF file includes:**

    Materials and Methods
    Supplementary Text
    Figs. S1 to S8

## Materials and Methods

<u>Section 1.1 Sample fabrication</u>

Mechanical exfoliation onto 90nm SiO$_2$/Si substrates was used to obtain ~20nm thick hBN and few-layer graphite flakes. hBN thickness was confirmed by atomic force microscopy. Standard polymer-based dry transfer techniques were used to form an hBN/graphite gate structure. Electrical contacts of electron beam evaporated Cr/Au (7/70 nm) were formed on the back gate structure via standard electron beam lithography. Contact mode AFM was used to clean polymer residue from the sample area of the backgate. 2H MoTe$_2$ (HQ Graphene) was mechanically exfoliated onto 285nm SiO$_2$/Si substrates in an argon glovebox, with H$_2$O and O$_2$ concentrations <0.1ppm. Monolayer flakes were identified via optical contrast. The complete dual gated structure was assembled using dry transfer within the inert glovebox environment using the "tear and stack" technique. The targeted twist angle is controlled by a high-precision rotation stage (Thorlabs PR01) with a typical accuracy of ~0.2°. The flakes were assembled as follows: top hBN, top gate graphite, top gate hBN, graphite grounding pin, tear-and-stack monolayer MoTe$_2$. The stack was then placed down on the prepared and cleaned backgate, such that the top gate graphite and grounding pin were electrically connected to the deposited contacts. The heterostructure geometry allows for full hBN encapsulation of the air-sensitive MoTe$_2$. The stamp polymer was dissolved in anhydrous chloroform for 5 minutes in a glovebox environment.

<u>Section 1.2 Optical measurements</u>

Photoluminescence and differential reflectance measurements were performed in reflection geometry in a home-built confocal optical microscope system. The sample was mounted in an exchange-gas cooled cryostat (attoDRY 2100) with a 9T superconducting magnet in Faraday geometry. A 632.8 nm HeNe laser and a Tungsten-Halogen lamp were used as excitation for the photoluminescence and reflectance measurements, respectively. The FWHM of the diffraction-limited excitation beam spot was ~1 µm. Photoluminescence and reflectance signals were dispersed with a 600 grooves/mm diffraction grating blazed at 1µm and detected using a LN cooled InGaAs CCD (Princeton Instruments PyLoN-IR 1.7). A long pass filter was used to remove laser from the photoluminescence signal before entrance into the spectrometer.

For the RMCD measurement, broad band white light generated by a SuperK Extreme Supercontinuum laser source was dual passed through a monochromator to select out a ~30meV band near the moiré trion resonance (see fig. S3). The filtered excitation was first passed through an optical chopper at 850Hz, and then through a 45° linear polarizer and photoelastic modulator (PEM) with a maximum retardance of $\lambda/4$. The PEM imparts a sinusoidal phase modulation at 50.1 kHz, allowing for excitation of the sample with alternating left and right circularly polarized light at this frequency. The signal reflected back from the sample was separated from the incidence path using a nonpolarizing beamsplitter and sent into an InGaAs avalanche photodiode. The photodiode signal was fed into two lock-in amplifiers tuned to 50.1 kHz and 850Hz, which detect the RMCD signal and the laser excitation intensity, respectively. Normalizing the RMCD signal by the laser intensity allows for the correction of any excitation power fluctuations.

<u>Section 1.3 Estimation of doping density n and displacement field D</u>

A parallel-plate capacitor model is used to calculate the gate induced carrier density *n* and displacement field *D* based on the applied gate voltages. The gate capacitances per area of the top and bottom gates, C$_{t/b}$= ε$_{hBN}$ε$_0$/$d_{t/b}$, are determined by the hBN thickness $d_{t/b}$ and dielectric constant, ε$_{hBN}$ = 3. The sample doping density *n* can be calculated as $n = (C_tV_t + C_bV_b)$, where $V_t$ and $V_b$ are the voltages applied to the top and bottom gates. The displacement field is defined as $D = (C_tV_t - $ $C_b V_b)/2\varepsilon_0 - D_{\text{offset}}$, where $\varepsilon_0$ is the vacuum permittivity and the subtracted value is the built-in offset. The offset was extracted from the mirror symmetry axis of the dual gate RMCD plot (Fig. 1E) and has a value of $D_{\text{offset}}$=0.04V/nm. The prominent spectral features appearing in the doping-dependent photoluminescence and differential reflectance spectra are used to assign the filling factors. From the doping density obtained from the capacitor model we can estimate the device twist angle, which is comparable to the targeted angle.

**Supplementary Text**

Section 2.1 Ground state calculations

The ground state calculations are performed using Hartree-Fock mean-field theory. The Hamiltonian is

$$H_{HF} = H_0 + H_H + H_F$$

where $H_0$ is the continuum model of twisted homobilayer MoTe$_2$ *(13, 14)*.

$$H_0 = \begin{pmatrix} h_k^t + E_D & w \\ w^\dagger & h_k^b \end{pmatrix},$$

where $h_k^\alpha = -\frac{(k - \kappa^\alpha)^2}{2m_\alpha} + V_k^\alpha, \alpha = t, b$ is the layer index, and $E_D$ is the interlayer bias. $V_\alpha(\boldsymbol{r}) = -2V_0 \sum_{i=1,3,5} \cos(\boldsymbol{g}_i \cdot \boldsymbol{r} + \phi_\alpha)$, where $\phi_t = -\phi_b = 89.6°$ is the layer dependent potential induced by the moiré *(13, 14)*, $\boldsymbol{g}_i$ is the first-shell reciprocal lattice vector, and $w(r) = w_0\left(1 + e^{i g_2 \cdot \boldsymbol{r}} + e^{i g_3 \cdot \boldsymbol{r}}\right)$ is the interlayer tunneling. We use $m_t = m_b = -0.62 m_e$, with $m_e$ the free-electron mass, $V_0 = 20$ meV, $w_0 = -8.5$ meV for all calculations. The calculations in Fig. 4E use an interlayer bias $E_D = 15$ meV.

In the continuum model, each Bloch wave function can be expanded in the plane-wave basis,

$$|k, m, \mu\rangle = \sum_{\alpha G} \varphi_{mk}(k + G, \alpha, \mu)|k + G, \alpha, \mu\rangle,$$

where $k$ is restricted to the first moiré Brillouin zone, $G$ is the reciprocal lattice vector, and $m$ and $\mu$ are the band and spin (valley) indices, respectively. The plane-wave spatial periodicity is consistent with the moiré pattern. In this plane-wave basis, Hartree and Fock self-energies *(41)* read

$$H_H\left(k + G, \alpha, \mu; k + G', \beta, \nu\right) = \left\langle k + G, \alpha, \mu \middle| \Sigma^H \middle| k + G', \beta, \nu\right\rangle$$

$$= \delta_{\alpha\beta} \delta_{\mu\nu} \frac{V_{\alpha\alpha'}\left(G - G'\right)}{\Omega}$$

$$\times \sum_{k'} \sum_{G''} \sum_{\alpha'\mu'} \delta\rho_{k'}\left(k + G' + G'', \alpha'\mu', k + G + G'', \alpha'\mu'\right)$$

and

$$H_F\left(k + G, \alpha, \mu; k + G', \beta, \nu\right) = \left\langle k + G, \alpha, \mu \middle| \Sigma^F \middle| k + G', \beta, \nu\right\rangle$$

$$= -\delta_{\mu\nu} \sum_{k'} \sum_{G''} \frac{V_{\alpha\alpha'}\left(k - k' - G''\right)}{\Omega} \delta\rho_{k'}\left(k' + G' + G'', \beta, \nu; k' + G + G'', \alpha, \mu\right),$$

where $\Omega$ is the system size and $\delta\rho = \rho - \rho_{iso}$ is the density matrix relative to an isolated twisted bilayer MoTe$_2$ filled to charge neutrality *(41)*. The density matrix is defined as

$$\rho_k\left(k+G,\alpha,\mu;k+G^{\,\prime},\beta,\nu\right)=\sum_m^{occupied}\varphi_{mk}^*(k+G,\alpha,\mu)\varphi_{mk}\left(k+G^{\,\prime},\beta,\nu\right).$$

and $V$ is the Coulomb coefficient, defined as

$$V_{\alpha\beta}(q)=\begin{cases}\dfrac{2\pi e^2}{\epsilon q},\alpha=\beta\\[2mm]\dfrac{2\pi e^2}{\epsilon q}e^{-qd},\alpha\neq\beta\end{cases},$$

where $\epsilon$ is the effective dielectric constant and $d$ is the effective layer distance between the two MoTe$_2$ single layers. We use $\epsilon=6$ and $d=7.3\text{Å}$ in all calculations.

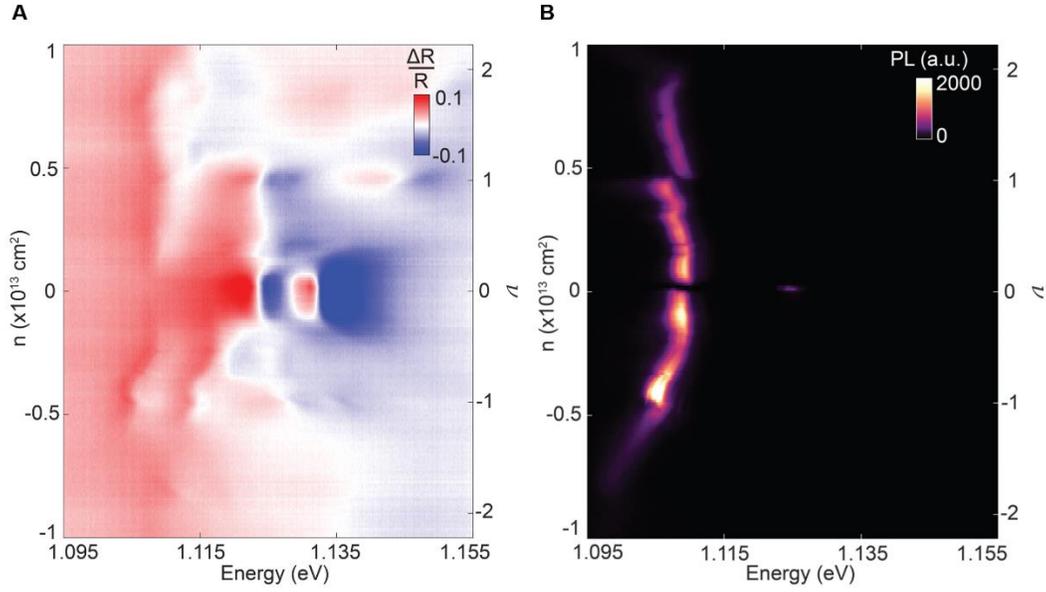

**Fig. S1. Filling factor assignment. (A)** Differential optical reflectance measurement, $\Delta R/R=(R_S-R)/R$, as a function of doping. $R_S$ (R) is the optical reflectance on the sample (off the sample). The assigned filling factors are denoted on the right axis. **(B)** Photoluminescence intensity plot as a function of doping. Both sets of data are taken under the same conditions. The pronounced spectral features in the optical reflectance plot match the trion photoluminescence reduction at particular doping levels, which are assigned as integer fillings of the moiré unit cell.

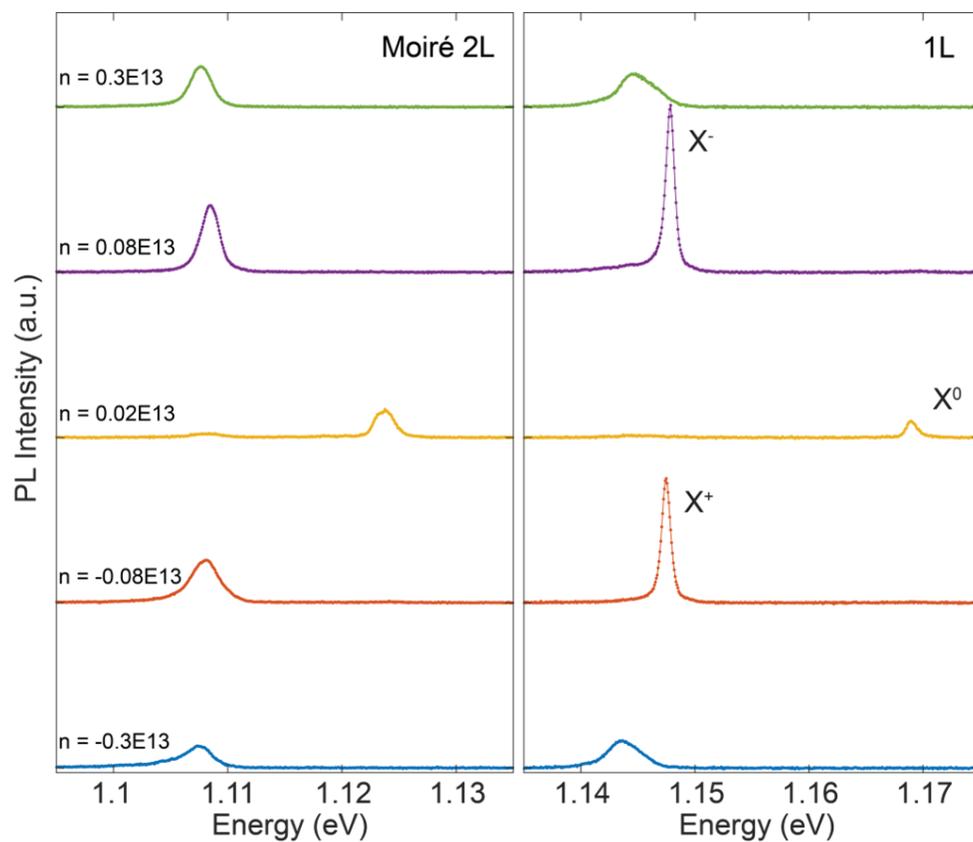

**Fig. S2. Comparison of photoluminescence spectra between monolayer (right panel) and twisted bilayer (left panel) MoTe₂ at selected doping.** Both sets of data are taken under the same conditions.

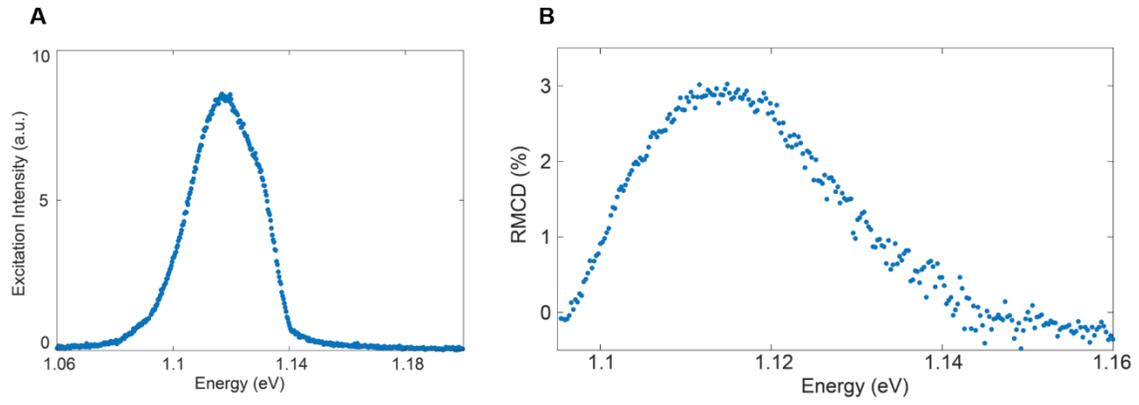

**Fig. S3. Photon energy dependent reflective magnetic circular dichroism (RMCD). (A)** Spectrum of the incident excitation light. **(B)** RMCD signal as a function of center energy of the photo-excitation at a magnetic field of 0.25T, $v$=-0.9, and $D$=-0.04V/nm.

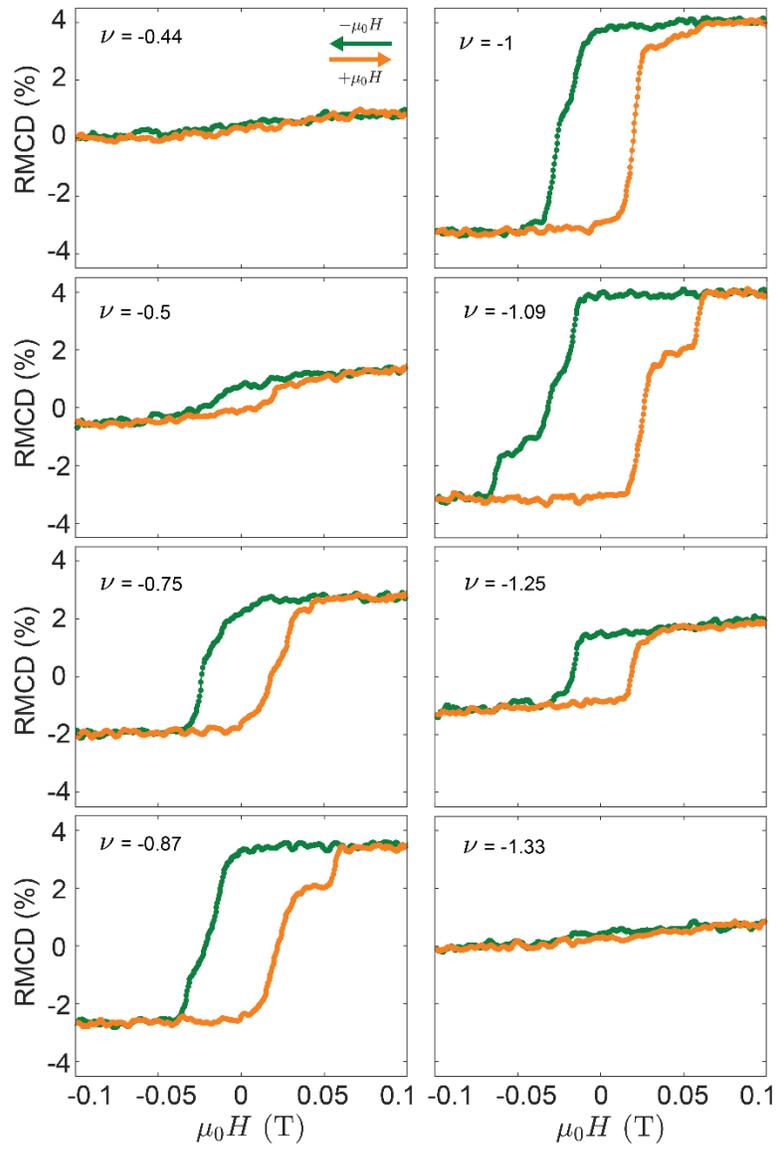

**Fig. S4. Filling factor dependent RMCD signal at *D*=0 V/nm.** RMCD signal as a function of out-of-plane magnetic field $\mu_0 H$ swept down and up at selected filling factors *v* indicated in each panel. The data are linecuts of the color plots of Fig. 2E in the main text. Magnetic domain effects near the spin flip transitions are apparent at certain filling factors (e.g. *v*=-0.87, -1.09).

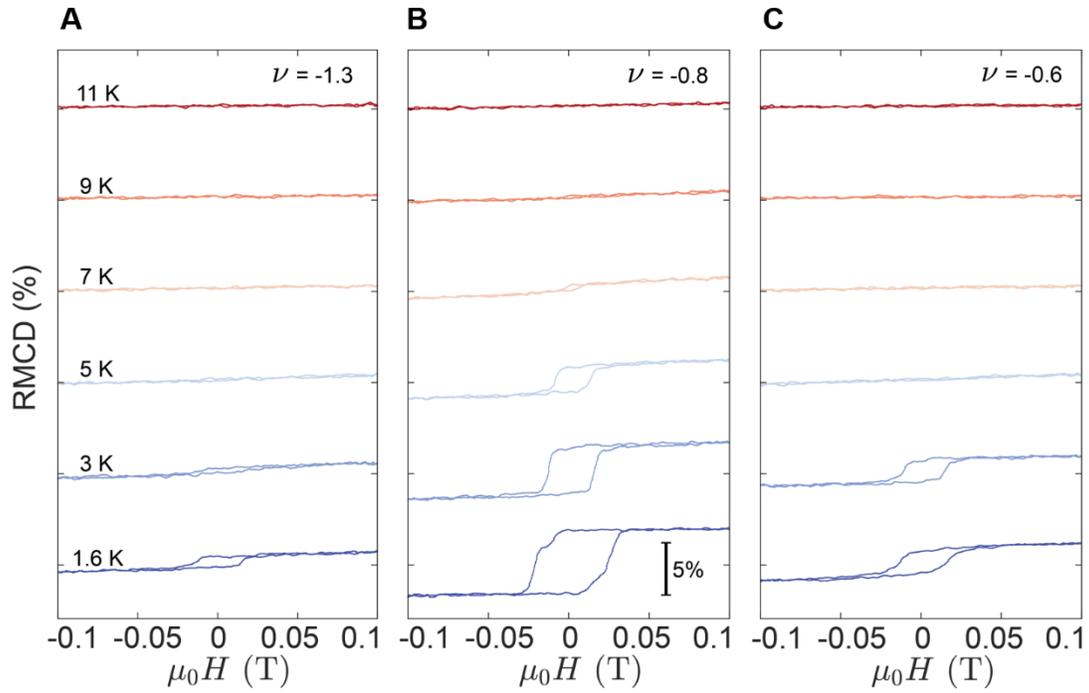

**Fig. S5. Temperature dependent RMCD signal at selected filling factors away from $v$=-1.** All data are taken at $D$=0 V/nm. **(A) (B)** and **(C)** correspond to filling factors $v$=-1.3, -0.8, and -0.6, respectively.

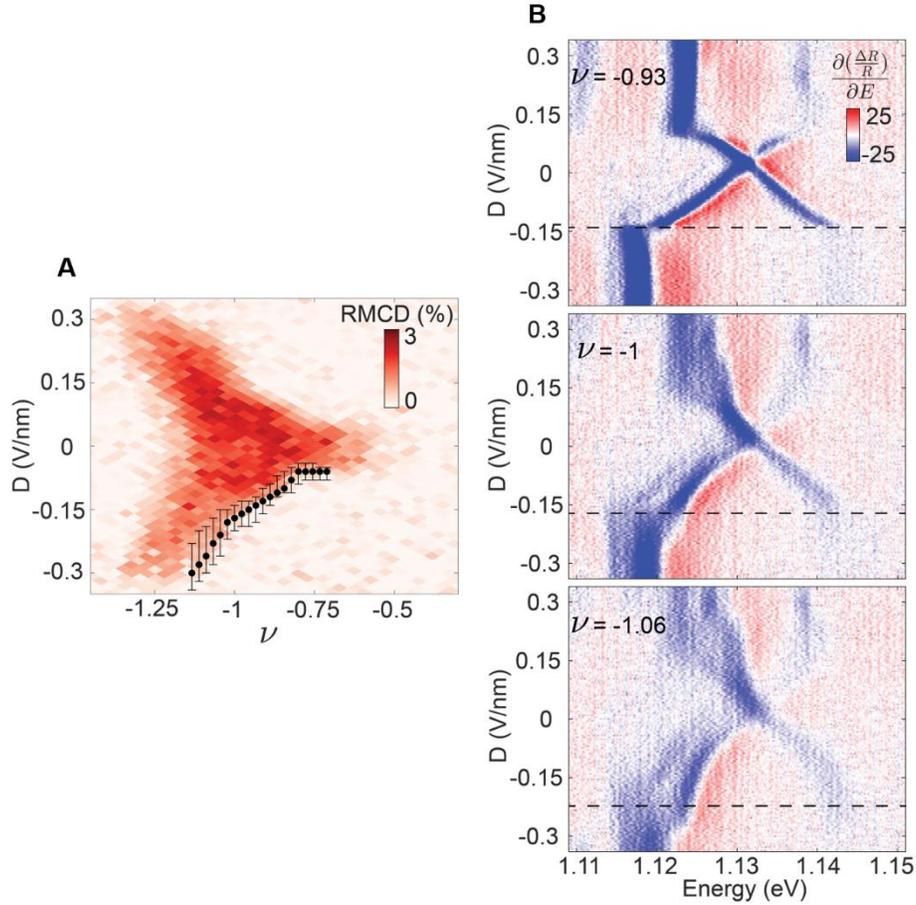

**Fig. S6. Full layer polarization assignment.** (**A**) RMCD signal intensity plot as a function of displacement field *D* and filling factor *v* (extracted from Fig. 1E in the main text). The black dots are the displacement field value for full layer polarization, determined by the approach shown in (B). Error bars correspond to the uncertainty in the D value of the transition extracted from reflectance data. (**B**) The intensity plot of the derivative of optical reflectance with respect to energy as a function of displacement field *D* at selected filling factors. The cross pattern versus *D* in the low *D* range is from the Stark shift of interlayer excitons. When full layer polarization is achieved, the feature ceases to shift with D. The critical D value is indicated by the dashed lines.

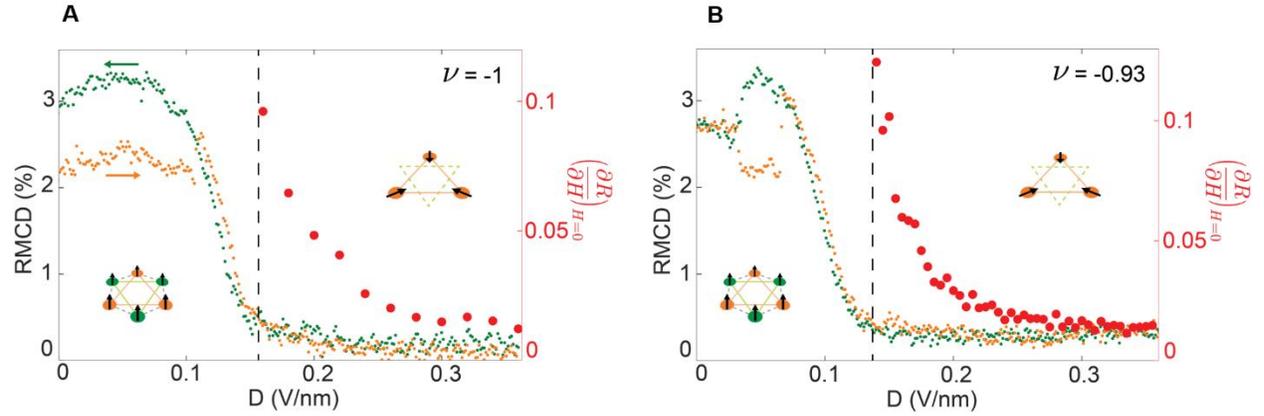

**Fig. S7. Displacement field control of magnetic interactions. (A)** RCMD signal at zero magnetic field is measured as $D$ is swept down (green) and up (orange) at $v$=-1. Red dots are the extracted RMCD slope $(\frac{\partial R}{\partial H})_{H=0}$ (arbitrary units) as $D$ approaches the ferromagnetic state from the phase space of antiferromagnetic interactions, showing singular behavior. Insets denote the spin arrangement in the triangular and honeycomb lattices, respectively. The different RMCD signal between different D sweeping directions away from the phase transition boundary is due to magnetic domain effects. **(B)** Same as (A) but at a slightly underdoped condition ($v$=-0.93), which shows similar phase transition behavior but with suppressed magnetic domain effects.

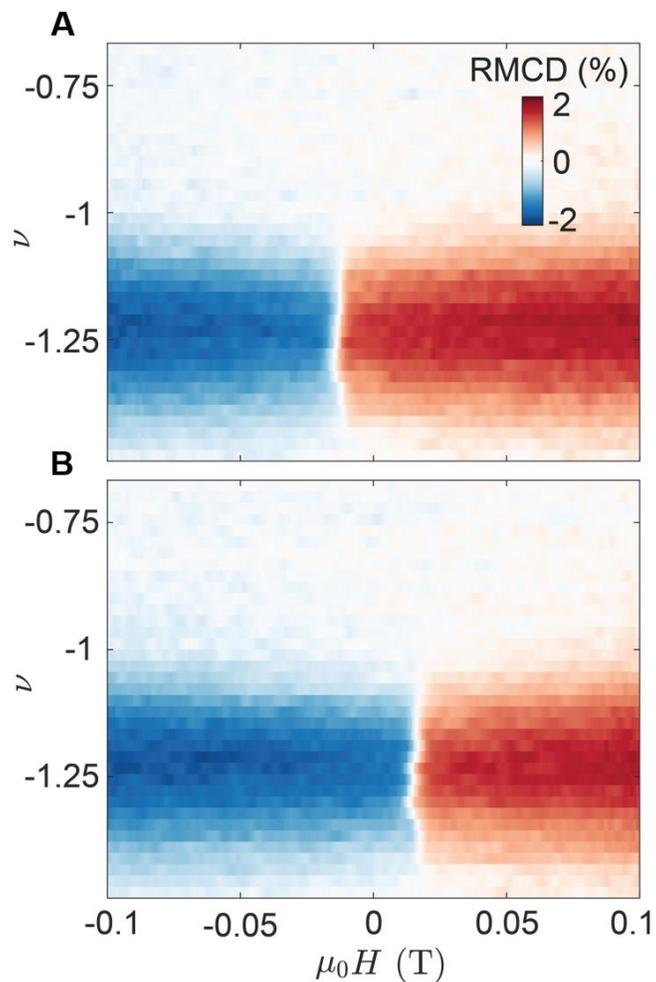

**Fig. S8. RMCD signal intensity plot versus magnetic field and filling factor.** All data are taken at $D$=0.26 V/nm. (**A**) and (**B**) correspond to magnetic field swept down and up, respectively.